\documentclass[pra,preprintnumbers,amsmath,amssymb,twocolumn,superscriptaddress]{revtex4}

\usepackage{graphicx}
\usepackage[utf8]{inputenc}

\graphicspath{ {figures/} }

\newcommand{\bra}[1]{\langle #1 |}

\newcommand{\ket}[1]{| #1 \rangle}
\newcommand{\braket}[2]{\langle #1| #2 \rangle}

\newcommand{\dyad}[2]{\ket{#1}\bra{#2}}

\newcommand{\proj}[1]{\ket{#1}\bra{#1}}

\newcommand{\tr}{\operatorname{tr}}

\newcommand{\ZZ}{\cc{z}}

\newcommand{\ee}{\vec{e}}
\newcommand{\kk}{\vec{k}}

\newcommand{\cc}[1]{{#1}^*}
\newcommand{\op}[1]{#1}

\renewcommand{\Re}{\mathrm{Re}}
\renewcommand{\Im}{\mathrm{Im}}

\newcommand{\opP}[1]{\ket{\pi_{#1}}\bra{\pi_{#1}}}
\newcommand{\bath}{\rm env}

\begin{document}
\title{Non-Markovian Quantum State Diffusion for Temperature-Dependent Linear Spectra of Light Harvesting Aggregates}

\author{Gerhard Ritschel}
\affiliation{Max-Planck-Institut f\"ur Physik komplexer Systeme, N\"othnitzer Str.\ 38,
D-01187 Dresden, Germany }

\author{Daniel Suess}
\affiliation{Institut f\"ur Theoretische Physik, Technische Universit\"at Dresden,
D-01062 Dresden, Germany}

\author{Sebastian M\"obius}
\affiliation{Max-Planck-Institut f\"ur Physik komplexer Systeme, N\"othnitzer Str.\ 38,
D-01187 Dresden, Germany }

\author{Walter T.\ Strunz}
\affiliation{Institut f\"ur Theoretische Physik, Technische Universit\"at Dresden,
D-01062 Dresden, Germany}

\author{Alexander Eisfeld}
\email{eisfeld@mpipks-dresden.mpg.de}
\affiliation{Max-Planck-Institut f\"ur Physik komplexer Systeme, N\"othnitzer Str.\ 38,
D-01187 Dresden, Germany }

\begin{abstract}

Non-Markovian Quantum State Diffusion (NMQSD) has turned out to be an efficient method to calculate excitonic properties of aggregates composed of organic chromophores, taking into account the coupling of electronic transitions to vibrational modes of the chromophores.
NMQSD is an open quantum system approach that incorporates environmental degrees of freedom (the vibrations in our case) in a stochastic way.
We show in this paper that for linear optical spectra (absorption, circular dichroism) \emph{no stochastics} is needed, even for finite temperatures.
Thus, the spectra can be obtained by propagating a {\em single} trajectory.
To this end we map a finite temperature environment to the zero temperature case using the so-called thermofield method.
The resulting equations can then be solved efficiently by standard integrators.
\end{abstract}

\maketitle

\section{Introduction}

Linear optical spectroscopy is an important tool to obtain information about multichromophoric complexes like J-aggregates \cite{Sc38_1_,BaKn00_66_,Ei07_321_,Sp09_4267_,SpDaeOu00_8664_,FiSeEn08_12858_} or photosynthetic light harvesting complexes \cite{GrHoVa91_194_,AmVaGr00__,MiBrGr10_257_,AdRe06_2778_}.
In particular the combination of linearly polarized spectra and circular dichroism spectra (difference of absorption of left and right circularly polarized light) allows one to draw many conclusions about the (often unknown) arrangement of the chromophores \cite{EiSeEn08_186_,SpDaeOu00_8664_} or to extract unknown electronic transition energies (often termed site energies) of the chromophores \cite{MiBrGr10_257_,AdRe06_2778_}.

The interpretation of measured aggregate spectra is complicated because of often strong coupling of an electronic excitation to internal vibrations (i.e.~nuclear coordinates of the chromophores) and other environmental degrees of freedom.
Therefore, one has to be careful not to draw conclusions solely based on an electronic exciton theory \cite{EiSeEn08_186_}.

The influence of vibrations on molecular aggregates has been studied using various methods (see for example~\cite{WiMo60_872_,Foe65_93_,BrHe71_865_,He77_1795_,LuFr77_36_,FrSi81_1166_,ScFi84_269_,KReMa96_99_,DaKoKl02_031919_,BeDaKj02_5810_,YaFl02_163_,Ei07_321_,EiKnBr07_104904_,FiSeEn08_12858_,IsFl09_234111_,Sp09_4267_,PrChHu10_050404_,RoEiDv11_054907_}).

Recently we have adopted the so-called Non-Markovian Quantum State Diffusion (NMQSD) approach \cite{St96_25_,Di96_309_,DiSt97_569_}, which allows one to treat aggregates where electronic excitation couples to a structured spectral density (which can, e.g., describe damped vibrational modes of the chromophores, see Appendix~\ref{app:aggregate}).

In the present work we will use a ``bath'' of harmonic oscillators to describe both internal (nuclear) and external degrees of freedom on the same footing.
The modes of this ``bath'' we will generally call vibrational modes (or simply vibrations).

The NMQSD method  is based on an open system approach.
In the present work we will take the open system to consist of the electronic states of the chromophores, as we have done in Refs.~\cite{RoEiWo09_058301_,RoStEi10_5060_,RoStEi11_034902_,RiRoSt11_113034_,RiRoSt11_2912_}. Note that one can also make other choices for the system, e.g.\ including some vibrational modes explicitly \cite{RoStEi11_034902_}.
In the NMQSD approach a stochastic Schr\"odinger equation that lives only in the space of the system degrees of freedom.
Averaging over the stochastic trajectories allows in principle to obtain {\it exactly} the reduced density operator of the system and expectation values of operators in the system space.

While for transfer of excitation the inclusion of stochastic terms in the NMQSD wave function equation is fundamental
\cite{RoEiWo09_058301_,RiRoSt11_113034_}, it turned out that for absorption {\it from the `total ground state'} (i.e.\ zero temperature) the stochastic noise terms do not enter explicitly the relevant propagation \cite{RoEiWo09_058301_,RoStEi11_034902_} and only a \emph{single} trajectory (without noise) needs to be propagated.

In the present work, we will show that the same formulation can be found for  finite temperatures through a mapping to temperature zero, which was introduced for the NMQSD approach in Refs.~\cite{DiGiSt98_1699_,Yu04_062107_}.
We will adopt this treatment in our calculations of optical properties.

The paper is organized as follows:
In the following Section~\ref{mod_of_agg} the model Hamiltonian used to describe the aggregate is introduced. 
Then in Section~\ref{calculation_of_spectra} we present our method for the calculation of linear spectra. First the general definitions of absorption and circular dichroism (CD) are given, then the thermofield approach providing the mapping to temperature zero is briefly reviewed.
In the following Subsections~\ref{sec:abs_in_qsd} and~\ref{sec:treat_func_deriv} the calculation of spectra within the NMQSD approach is derived. 
In Section~\ref{sec:example_calculations} we apply the method to calculate absorption and CD spectra of a model dimer.

In several appendices we provide additional details.

Throughout the paper we set $\hbar = 1$ and $k_\mathrm{B} = 1$.

\section{The aggregate Hamiltonian}
\label{mod_of_agg}


We consider an aggregate consisting of $N$ monomers, labeled by $m = 1,\ldots,N$.
We assume that the electronic wave functions of the monomers do not overlap and take them to be real.
The electronic ground state of the aggregate, from which absorption takes place, is then taken as the product
\begin{equation}
\label{ElGroundStateAgg}
 \ket{g_{\rm el}}=\prod_{m=1}^N \ket{\phi^g_m}
\end{equation}
of the electronic ground states $\ket{\phi_m^g}$ of all individual monomers.
Since we are interested in linear optical properties we will restrict the excited state basis to states in which only one monomer $n$ is electronically excited and all other monomers are in their electronic ground state. These states are denoted by
\begin{equation}
\label{pi_n_states}
 \ket{\pi_n}=\ket{\phi^e_n}\prod_{m\ne n}^N \ket{\phi^g_m}.
\end{equation}
We expand the aggregate Hamiltonian w.\,r.\,t.\ the states defined in Eqs.~(\ref{ElGroundStateAgg}) and (\ref{pi_n_states}) and neglect states with more than one electronic excitation on the aggregate. 
We then write the Hamiltonian of the aggregate as
\begin{equation}
\label{eq:H_agg}
 H=H^g+H^e.
\end{equation}
Here, $H^g$ is the electronic ground state Hamiltonian (for a brief derivation see Appendix~\ref{app:aggregate})
\begin{equation}
H^g= H_{\rm env}\ket{g_{\rm el}}\bra{g_{\rm el}},
\end{equation}
where 
\begin{equation}
\label{eq:H_env}
H_{\rm env} =\sum_{n=1}^N\sum_{\lambda}\omega_{n\lambda}a^{\dagger}_{n\lambda}a_{n\lambda} 
\end{equation}
comprises all the (possibly different) vibrational modes $\lambda$ of the monomers and will later play the role of an environment.

The Hamiltonian in the one-excitation subspace is given by (see Appendix~\ref{app:aggregate})
\begin{equation}
\label{HeAsSysPlusIntPlusEnv}
  H^e=H_{\rm sys}+H_{\rm int}+H_{\bath},
\end{equation}
with the purely electronic ``system'' part
\begin{equation}
\label{H_sys}
  H_{\rm sys}=\sum_{n=1}^N\varepsilon_n\ket{\pi_n}\bra{\pi_n}+\sum_{n,m=1}^N V_{nm}\ket{\pi_n}\bra{\pi_m},
\end{equation}
where the matrix element $V_{nm}$ causes electronic excitation to be transferred from monomer $m$ to monomer $n$,
and a part $H_{\bath}$ describing the ``bath'' of vibrational modes given by Eq.~(\ref{eq:H_env}).

The coupling of electronic excitation to these vibrations is expressed through
\begin{equation}
\label{HInt}
  H_{\rm int} = \sum_{n=1}^N K_n \sum_{\lambda}\kappa_{n\lambda}(a^{\dagger}_{n\lambda}+a_{n\lambda})
\end{equation}
with the (Hermitian) system operators $K_n = -\proj{\pi_n}$.
Note that also the inclusion of off-diagonal coupling terms
\begin{equation}
  H_{\rm int}^{\rm off} = \sum_{n>m} K_{nm}\sum_{\lambda}\kappa_{nm\lambda}(a^{\dagger}_{nm\lambda}+a_{nm\lambda})
\end{equation}
with $K_{nm} = -\frac{1}{2}\left( \ket{\pi_n}\bra{\pi_m} + \ket{\pi_m}\bra{\pi_n} \right)$
is possible. The generalization of the following equations to this case is straightforward.
However, for readability we restrict ourselves to the diagonal coupling terms Eq.~(\ref{HInt})
and also assume that the baths are uncorrelated.
Note that in our previous work we used the symbol $L_n$ instead of $K_n$ for the system coupling operators. Here, we only use $L_n$ in Appendix~\ref{sec:general_equations_of_motion} in order to emphasize that the derivation therein also holds for coupling operators that are not necessarily Hermitian.

It is convenient to define the so-called spectral density
\cite{MaKue11__} of monomer $n$ by
\begin{equation}
\label{eq:spec_dens}
J_{n}(\omega) = \pi \sum_{\lambda}|\kappa_{n\lambda}|^2\ \delta(\omega-\omega_{n\lambda}).
\end{equation}
Later, $J(\omega)$ will be considered as a continuous function of frequency.
For an interpretation of the spectral density see Appendix~\ref{app:aggregate}.

\section{Calculation of linear optical spectra}
\label{calculation_of_spectra}
\subsection{Transition dipole operator}
The transition dipole operator $\hat{\vec{\mu}}_n$ of monomer $n$ is assumed to be independent of nuclear (environmental) coordinates and is written as
\begin{equation} 
 \hat{\vec{\mu}}_n = \vec{\mu}_n(\dyad{\phi^g_n}{\phi^e_n} + \dyad{\phi^e_n}{\phi^g_n}).
\end{equation}
Here, $\vec{\mu}_n$ denotes the transition dipole moment of monomer $n$ (see, e.g., Ref.~\cite{MaKue11__}).
The  dipole operator of the aggregate is given by the sum 
\begin{equation}
\label{eq:dip_op_agg}
 \hat{\vec\mu} = \sum_{n} \hat{\vec\mu}_n.
\end{equation}
This is the basic quantity that enters the calculation of optical spectra.

\subsection{Absorption and circular dichroism}
The transition strength for the linear optical spectra can be obtained from a half-sided
\footnote{If one extends the dipole-auto-correlation function $c(t)$ to negative times such that $c(-t) = c^*(t)$, the transition strength can be calculated as
$
 F(\omega) = \int_{-\infty}^{\infty} dt \, e^{i\omega t}\, c(t).
$}
Fourier transformation \cite{Mu95__, MaKue11__}
\begin{equation}
\label{eq:spectrum}
 F(\omega) = \mathrm{Re} \; \int_0^{\infty} dt \, e^{i \omega t}\, c(t).
\end{equation}
The explicit form of the correlation function $c(t)$ will be specified below for the case of absorption and CD. The total initial state of system and bath prior to light absorption is taken as
\begin{equation}
\label{eq:rho_0}
 \rho_0=\ket{g_{\rm el}}\bra{g_{\rm el}}\otimes \rho_{\bath},
\end{equation}
where the aggregate is in its electronic ground state Eq.~(\ref{ElGroundStateAgg}) and the bath is in a thermal state at temperature $T$, i.e.\ 
\begin{equation}
\label{eq:rho_th}
 \rho_{\bath} \equiv \rho(\beta) = \frac{ e^{-\beta H_{\bath}} } { \tr_{\bath} \{ e^{-\beta H_{\bath}} \} }
\end{equation}
with the inverse temperature $\beta = 1 / T$.
In Eq.~(\ref{eq:rho_th}), $\tr_{\bath}$ denotes the trace over the bath degrees of freedom.
Since for the molecules we have in mind (e.g.~chlorophyll or cyanine dyes) the electronic transition energies are on the order of an eV, one can safely ignore thermally excited electronic states, as has been done in Eq.~(\ref{eq:rho_0}).

In order to calculate the correlation function $c(t)$ in Eq.~(\ref{eq:spectrum}) we introduce a correlation operator $\mathcal{C}(t)$ in the electronic system
defined by
\begin{equation}
\label{eq:def:C_nm}
 \mathcal{C}(t)\equiv \tr_{\bath} \{ (U^g)^{\dagger}(t)\, U^e(t)\, \rho_{\bath} \}
\end{equation}
with the propagators for system and bath,
\begin{equation}
\begin{split}
\label{eq:def:U}
U^g(t) &\equiv e^{-iH^g t}, \\
U^e(t) &\equiv  e^{-iH^e t}, 
\end{split}
\end{equation}
in the respective electronic states.
In linear response theory, one can then calculate $c(t)$ in Eq.~(\ref{eq:spectrum}) for the absorption spectrum of an isotropically oriented sample as
\begin{align}
\label{eq:def:C_abs}
 c^{\rm Abs}(t) =& \tr_{\rm sys} \{ A^{\rm Abs}\, \mathcal{C}(t) \} \\
=&\sum_{nm}  (A^{\rm Abs})_{nm}\, \mathcal{C}_{nm}(t),
\end{align}
where $A^{\rm Abs}$ is a matrix in the electronic system with matrix elements
\begin{equation}
\label{eq:A_abs}
 (A^{\rm Abs})_{nm} = \bra{\pi_n}A^{\rm Abs}\ket{\pi_m} = \vec{\mu}_n\cdot \vec{\mu}_m
\end{equation}
and the dot denotes the real scalar product.
Similarly, for the CD spectrum within the Rosenfeld formalism \cite{Ro29_161_,Ki37_479_} one has
\begin{equation}
\label{eq:def:C_CD}
c^{\rm CD}(t) = \tr_{\rm sys} \{ A^{\rm CD}\, \mathcal{C}(t) \},
\end{equation}
where the matrix elements of $A^{\rm CD}$ are given by
\begin{equation}
\label{eq:A_CD}
 (A^{\rm CD})_{nm} = \bra{\pi_n}A^{\rm CD}\ket{\pi_m} = \vec{R}_{nm}\cdot (\vec{\mu}_n\times \vec{\mu}_m).
\end{equation}
Here, $\vec{R}_{nm}=\vec{R}_m-\vec{R}_n$ denotes the distance vector between monomers $n$ and $m$.
The validity of the Rosenfeld approximation and a more appropriate formula for large aggregates with considerable excitonic delocalization are discussed in \cite{EiKnBr07_104904_}. In Appendix~\ref{sec:c(t)} short derivations of $c^{\rm Abs}(t)$ and $c^{\rm CD}(t)$ can be found.

\subsection{Thermofield method}

In order to evaluate the expression $\mathcal{C}(t)$
defined in Eq.~(\ref{eq:def:C_nm})
we use the thermofield approach \cite{SeUm83_196_}: The thermal initial state $\rho(\beta)$ of the environment is mapped onto the (pure) ground state $\ket{0(\beta)}$ (later denoted as ``thermal vacuum'') of an enlarged environment with suitably constructed creation and annihilation operators, as detailed below.
The calculation of $\mathcal{C}(t)$, which will be discussed in the following Section~\ref{sec:abs_in_qsd}, then proceeds analogously to the temperature zero case~\cite{RoEiWo09_058301_,RoStEi11_034902_}.
In this subsection we sketch briefly the procedure for the mapping to the enlarged bath, following the treatment of Refs.~\cite{DiGiSt98_1699_,Yu04_062107_}:

First, in addition to the physical bath operators $(a_{n\lambda}, a^{\dagger}_{n\lambda})$ one introduces independent `fictitious' negative frequency bath operators $(b_{n\lambda}, b^{\dagger}_{n\lambda})$ resulting in the expression
\begin{equation}
\label{eq:def:H_env_thermofield}
 \bar{H}_{\bath} \equiv H_{\bath} + H_b
\end{equation}
for the new environmental Hamiltonian, with
\begin{equation}
\label{eq:def:H_b}
 H_b = \sum_{n=1}^N\sum_{\lambda}(-\omega_{n\lambda}) b^{\dagger}_{n\lambda}b_{n\lambda}
\end{equation}
and the same (negative) frequencies $\omega_{n\lambda}$ as for the physical bath $H_{\bath}$.
With the bar, as in Eq.~(\ref{eq:def:H_env_thermofield}), we indicate that the Hamiltonian contains the additional $b$-bath.
As a consequence, the number of degrees of freedom of the new bath Hamiltonian
\begin{equation}
\label{eq:H_bath_doubled}
 \bar{H}_{\bath} = \sum_{n=1}^N\sum_{\lambda}\omega_{n\lambda} \big( a^{\dagger}_{n\lambda}a_{n\lambda} - b^{\dagger}_{n\lambda}b_{n\lambda} \big)
\end{equation}
is twice that of the original one. 
The desired state $\ket{0(\beta)}$ in the doubled-bath Hilbert space is now constructed such that one recovers the correct thermal equilibrium state $\rho(\beta)$ defined in Eq.~(\ref{eq:rho_th}) for the physical $a$ bath after tracing out the fictitious $b$ degrees of freedom, i.e.\ 
\begin{equation}
\label{eq:thermal_vacuum}
 \rho(\beta) = \tr_{b} \ket{0(\beta)}\bra{0(\beta)}.
\end{equation}
Note that the additional degrees of freedom do not alter the dynamics, because they are uncoupled from the physical ones.
Through a (temperature-dependent) Bogoliubov transformation of the bath operators $(a_{n\lambda}, a^\dagger_{n\lambda})$ and $(b_{n\lambda}, b^\dagger_{n\lambda})$ one defines so-called thermal bath annihilation operators
\begin{equation}
\label{eq:thermal_operators}
 \begin{split}
 A_{n\lambda} = & \sqrt{\bar{n}_{n\lambda}+1} \,a_{n\lambda} - \sqrt{\bar{n}_{n\lambda}} \,b^\dagger_{n\lambda} \\
 B_{n\lambda} = & \sqrt{\bar{n}_{n\lambda}+1} \,b_{n\lambda} - \sqrt{\bar{n}_{n\lambda}} \,a^\dagger_{n\lambda}
 \end{split}
\end{equation}
and their corresponding adjoint creation operators. Here, $\bar{n}_{n\lambda} = \left( {e^{\beta\omega_{n\lambda}}-1} \right)^{-1}$ is the mean thermal occupation number of the physical mode $\lambda$ of monomer $n$.
The thermal operators fulfill the same bosonic commutation relations as the original $a$ and $b$ operators and thus create their own harmonic oscillator algebra.
As a result, the thermal annihilation operators $A_{n\lambda}$ and $B_{n\lambda}$ annihilate the thermal vacuum $\ket{0(\beta)}\equiv\ket{0}_A \ket{0}_B$.
Here $\ket{0}_A$ is a shorthand notation for the product vector of all environmental $(A_{n\lambda}, A^\dagger_{n\lambda})$ oscillators being in their vacuum state $\ket{0_{n\lambda}}_A$ respectively, i.e. $\ket{0}_A = \prod_{n\lambda} \ket{0_{n\lambda}}_A$ (and the analogous expression for $\ket{0}_B$).
The extended environmental Hamiltonian with the untransformed bath operators expressed in terms of the thermal bath operators now reads
\begin{equation}
\label{eq:bar_H_g}
 \bar{H}_{\bath} = \sum_{n=1}^N\sum_{\lambda}\omega_{n\lambda} \big( A^{\dagger}_{n\lambda}A_{n\lambda} - B^{\dagger}_{n\lambda}B_{n\lambda} \big).
\end{equation}
For the electronic ground state Hamiltonian one thus obtains
\begin{equation}
\label{eq:bar_H_g}
 \bar{H}^g = \bar{H}_{\bath} \proj{g_{\rm el}},
\end{equation}
whereas for the electronically excited state one gets
\begin{equation}
\label{eq:bar_H_e}
 \bar{H}^e = H_{\rm sys} + \bar{H}_{\bath} + H_{\rm int}
\end{equation}
with
\begin{equation}
\begin{split}
 H_{\rm int} = & \sum_{n=1}^N K_n \sum_{\lambda}\kappa_{n\lambda} \times \\
 & \Big( \sqrt{\bar{n}_{n\lambda}+1}\big(A^{\dagger}_{n\lambda}+A_{n\lambda}\big)
 + \sqrt{\bar{n}_{n\lambda}}\big(B^{\dagger}_{n\lambda}+B_{n\lambda}\big) \Big).
\end{split}
\end{equation}

\subsection{Absorption in the NMQSD approach}
\label{sec:abs_in_qsd}

To evaluate the expression Eq.~(\ref{eq:def:C_nm}) for the correlation operator $\mathcal{C}(t)$ we apply the thermofield approach outlined in the previous section and insert $\rho(\beta)$ from Eq.~(\ref{eq:thermal_vacuum}) into Eq.~(\ref{eq:def:C_nm}) to obtain
\begin{align}
 \mathcal{C}(t) & = \tr_a \tr_b \big\lbrace (U^g)^{\dagger}(t)\, U^e(t)\, \proj{0(\beta)} \big\rbrace.
\end{align}
Here, $\tr_a = \tr_{\bath}$ denotes the trace over the original environmental degrees of freedom and $\tr_b$ is, as in Eq.~(\ref{eq:thermal_vacuum}), the trace over the negative frequency oscillators.
Inserting $\openone = e^{iH_b t}e^{-iH_b t}$, where $H_b$ is the Hamiltonian for the additional bath degrees of freedom defined in Eq.~(\ref{eq:def:H_b}), between the two propagators and rearranging the trace yields
\begin{align}
 \mathcal{C}(t) & = \tr_a \tr_b \big\lbrace \bar{U}^e(t)\, \proj{0(\beta)}\, (\bar{U}^g)^{\dagger}(t) \big\rbrace.
\end{align} 
Analogous to Eq.~(\ref{eq:def:U}), the propagators are defined as $\bar{U}^{g}(t) = e^{-i\bar{H}^g t}$ and $\bar{U}^{e}(t) = e^{-i\bar{H}^e t}$ with $\bar{H}^g$ and $\bar{H}^e$ given by Eqs.~(\ref{eq:bar_H_g}) and (\ref{eq:bar_H_e}).
Note that the trace is over the full doubled-bath Hilbert space. 
Since $\bra{0(\beta)}(\bar{U}^g)^{\dagger}(t)=\bra{0(\beta)}$, performing the trace leads to   
\begin{align}
\label{eq:CorrelationOperatorThermofield}
 \mathcal{C}(t) & = \bra{0(\beta)} \bar{U}^{e}(t) \ket{0(\beta)}.
\end{align}

The aim in the following is to derive an evolution equation for $\mathcal{C}(t)$ that can be handled numerically in an efficient way. To this end we expand the environmental part of the Hilbert space in terms of Bargmann coherent states~\cite{Ba61_187_}
$\ket{\zeta_{n\lambda}}_A = \exp{(\zeta_{n\lambda}A^{\dagger}_{n\lambda})}\ket{0_{n\lambda}}_A$ and
$\ket{\xi_{n\lambda}}_B = \exp{(\xi_{n\lambda}B^{\dagger}_{n\lambda})}\ket{0_{n\lambda}}_B$
 of the $A$ and $B$ vibrations. Here the $\zeta_{n\lambda}$ as well as the $\xi_{n\lambda}$ are complex numbers. 
Defining $\ket{{\boldsymbol{\zeta}}}_{A} = \prod_{n\lambda}\ket{\zeta_{n\lambda}}_{A}$, and $d^2\boldsymbol{\zeta}=d\, \Re(\boldsymbol{\zeta})\ d\,\Im(\boldsymbol{\zeta})$ (and the corresponding expressions for $\boldsymbol{\xi}$) the completeness relations for the Bargmann states are given by~\cite{Ba61_187_, Ga09__}
\begin{equation}
\label{eq:completeness_Bargmann}
\begin{split}
 \int\frac{d^2 \boldsymbol{\zeta}}{\pi}e^{-|\boldsymbol{\zeta}|^2} \ket{\boldsymbol{\zeta}}_{A} \bra{\boldsymbol{\zeta}}_{A} = \openone_A, \\
 \int\frac{d^2 \boldsymbol{\xi}}{\pi}e^{-|\boldsymbol{\xi}|^2} \ket{\boldsymbol{\xi}}_{B} \bra{\boldsymbol{\xi}}_{B} = \openone_B.
\end{split}
\end{equation}

Using this expansion, one can derive an evolution equation \cite{DiGiSt98_1699_} for a quantity from which $\mathcal{C}(t)$ can be obtained.
The general form of the resulting equations, which also hold for system coupling operators that are not Hermitian, is discussed in Appendix~\ref{sec:general_equations_of_motion}.
In the following, in order to simplify the notation, we will use the fact that for the present case where $K_n=K_n^{\dagger}$
the general equation (\ref{eq:EvolEqReducedPropagator}) can be written as (see Appendix~\ref{sec:general_equations_of_motion})
\begin{equation}
\label{EvolEqReducedPropagator}
\begin{split}
\partial_t G(t,\boldsymbol{z}^*) =
& -i H_{\rm sys}G(t,\boldsymbol{z}^*)
  +\sum_n K_n z^*_{t,n}G(t,\boldsymbol{z}^*) \\
& -\sum_n K_n \int_0^t ds\ \alpha_n(t-s)\frac{\delta}{\delta z^*_{s,n}} G(t,\boldsymbol{z}^*),
\end{split}
\end{equation}
which is the NMQSD equation \cite{DiGiSt98_1699_} in its linear form (and for Hermitian $K_n$)
\footnote{Using the Markov approximation $\alpha_n(t-s)=\gamma\delta(t-s)$ one obtains the well-known Markovian QSD equation \cite{GiPe92_5677_, Pe98__} (here in its linear form).},
with the reduced propagator
\begin{equation}
\label{eq:def:G_z}
 G(t,\boldsymbol{z}^*) \equiv \bra{\boldsymbol{z}} \bar{U}^{e}(t) \ket{0(\beta)}
\end{equation}
and the initial condition $G(0,\boldsymbol{z}^*) = \openone_{\rm sys}$.
In Eq.~(\ref{EvolEqReducedPropagator}) $\alpha_n (\tau)$ is the bath correlation function of monomer $n$ given by
\begin{equation}
\label{eq:def:alpha_with_lambda}
\begin{split}
\alpha_n(\tau) = \sum_{\lambda}|\kappa_{n\lambda}|^2 \Big( \coth\big(\frac{\omega_{n\lambda}}{2T}\big) \cos(\omega_{n\lambda} \tau)  -i\sin(\omega_{n\lambda} \tau) \Big)
\end{split}
\end{equation}
and the $z^*_{t,n}$ are defined as
\begin{equation}
z^*_{t,n} = -i \sum_{\mu} g_{n\mu} e^{i \tilde{\omega}_{n\mu} t} z^*_{n\mu}.
\end{equation}
The numbers $z_{n\mu}$ are defined in Appendix~\ref{sec:general_equations_of_motion} and combine $\xi_{n\mu}$ and $\zeta_{n\mu}$.
The factors $g_{n\mu}$ and frequencies $\tilde{\omega}_{n\mu}$ are also defined in Appendix~\ref{sec:general_equations_of_motion}.

Note, that the $z^*_{t,n}$ are deterministic time-dependent complex functions.
However, in Refs.~\cite{DiSt97_569_, DiGiSt98_1699_, RoEiWo09_058301_} they play the role of stochastic processes driving the evolution of the system---a viewpoint that is not necessary for this paper.  
Nonetheless, later on in this work we will also call $z^*_{t,n}$ a stochastic process.

It is important to note that the $z_{n\mu}^*$ have the same basic properties as the $\xi_{n\lambda}^{*}$ and $\zeta_{n\lambda}^{*}$, in particular $\int\frac{d^2 z_{n\mu}}{\pi}\,e^{-|z_{n\mu}|^2} (z^*_{n\mu})^j=\delta_{0j}$. 
The reduced propagator $G(t,\boldsymbol{z}^*)$ is analytic in $\boldsymbol{z}^*$ because of the properties of the Bargmann states \cite{Ba61_187_}
and we can Taylor expand it resulting in
\begin{equation}
\begin{split}
\label{eq:G_z_Taylor}
 G(t,\boldsymbol{z}^*) =
 & G^{(0)}(t)
   + \sum_{n_1\lambda_1} G^{(1)}_{n_1 \lambda_1}(t) z_{n_1 \lambda_1}^* \\
 & + \sum_{\stackrel{n_1 n_2}{\lambda_1 \lambda_2}} G^{(2)}_{\stackrel{n_1 n_2}{\lambda_1 \lambda_2}}(t) z_{n_1 \lambda_1}^* z_{n_2 \lambda_2}^* + \dots.
\end{split}
\end{equation}
Note that the $\boldsymbol{z}^*$-independent part $G^{(0)}(t)$ is the same as in Eq.~(\ref{eq:G_zeta_xi_Taylor}) and that Eq.~(\ref{eq:C(t)=G^{(0)}(t)}) still holds. Analogous to Eqs.~(\ref{eq:correlation_operator_with_G_of_zeta_xi}) and (\ref{eq:C(t)=G^{(0)}(t)}) we have
\begin{equation}
\label{eq:correlation_operator_with_G_of_z}
 \mathcal{C}(t) = \int \frac{d^2 \boldsymbol{z}}{\pi}\,e^{-|\boldsymbol{z}|^2}\,G(t,\boldsymbol{z}^*)
                = G^{(0)}(t).
\end{equation}

Using Eq.~(\ref{EvolEqReducedPropagator}), we will derive an evolution equation for $\mathcal{C}(t)$, in the following.
Equation~(\ref{EvolEqReducedPropagator}) is exact. It describes the full evolution of the electronic system in the excited state manifold coupled to the environment. Note, however, that the appearance of the functional derivative in Eq.~(\ref{EvolEqReducedPropagator}) renders a general solution very difficult, even though for the case of absorption considered here we are only interested in $G^{(0)}(t)$, which is independent of the complex numbers (stochastic processes) $z^*_{t,n}$. There are only a few cases where exact solutions to Eq.~(\ref{EvolEqReducedPropagator}) are known \cite{DiGiSt98_1699_,StDiGi99_1801_,StYu04_052115_,JiZhYo12_042106_}.
In general, it might not even be possible to derive an exact closed equation for $\mathcal{C}(t)=G^{(0)}(t)$, but we can still make use of the integrals over the environmental degrees of freedom in Eq.~(\ref{eq:correlation_operator_with_G_of_z}) in order to find a tractable expression for $\mathcal{C}(t)$ that does not involve the $z^*_{t,n}$ anymore.

To achieve this, we first note that using Eq.~(\ref{eq:correlation_operator_with_G_of_z}) we find from the evolution equation~(\ref{EvolEqReducedPropagator}) for the reduced propagator $G(t,\boldsymbol{z}^*)$ (for the considered case of self-adjoined coupling) the evolution equation
\begin{equation}
\label{EvolEqCorrelatorWithNoise}
\begin{split}
 \partial_t \mathcal{C}(t) = &\int \frac{d^2 \boldsymbol{z}}{\pi}\,e^{-|\boldsymbol{z}|^2} \times \Big( -i H_{\rm sys} + \sum_n K_n z^*_{t,n} \\
 &-\sum_n K_n \int_0^t ds\ \alpha_n(t-s)\frac{\delta}{\delta z^*_{s,n}}  \Big) G(t,\boldsymbol{z}^*)
\end{split}
\end{equation}
for the correlation operator $\mathcal{C}(t)$.
Using $\int\frac{d^2 \boldsymbol{z}}{\pi}\,e^{-|\boldsymbol{z}|^2} (z^*_{t,n})^{j} = \delta_{0j}$, the integrals in the first two terms can be performed and Eq.~(\ref{EvolEqCorrelatorWithNoise}) can be rewritten to
\begin{equation}
\label{EvolEqCorrelatorWithNoise2}
\begin{split}
 \partial_t \mathcal{C}(t) =
 -i H_{\rm sys}\, \mathcal{C}(t) -\sum_n K_n \int \frac{d^2 \boldsymbol{z}}{\pi}\,e^{-|\boldsymbol{z}|^2} \\
 \times \int_0^t ds\ \alpha_n(t-s)\frac{\delta}{\delta z^*_{s,n}} G(t,\boldsymbol{z}^*).
\end{split}
\end{equation}
The second term in Eq.~(\ref{EvolEqCorrelatorWithNoise}) vanishes, because $G(t,\boldsymbol{z}^*)$ is analytic in $\boldsymbol{z}^*$ and thus its product with $z^*_{t,n}$ is a sum of powers of the form $(z^*_{t,n})^j$.
The difficult third term still prevents us from having a closed equation for $\mathcal{C}(t)$.
To tackle this problem, we show an approach in Section~\ref{sec:treat_func_deriv}, which leads to a system of (infinitely many) coupled equations from which $\mathcal{C}(t)$ can be approximated numerically.
A second approach that can often be useful and has been employed in the past is outlined in Appendix~\ref{sec:ZOFE-approach}.

\subsection{Handling of the functional derivative}
\label{sec:treat_func_deriv}

\subsubsection{Hierarchy of pure states (HOPS):}

In our previous work \cite{RoEiWo09_058301_, RoStEi10_5060_, RoStEi11_034902_} we used an ansatz for the functional derivative $\frac{\delta}{\delta z_{s,n}^*}$. This ansatz is briefly discussed in Appendix~\ref{sec:ZOFE-approach}.
However, it is not {\it a priori} clear that this ansatz is always justified.

Recently, we developed a  scheme that does not use the abovementioned ansatz (see Eq.~(\ref{eq:def:Oop}))  to treat the complicated last term in Eq.~(\ref{EvolEqCorrelatorWithNoise}).
This scheme leads to an exact solution of the original problem that can be handled numerically in an efficient way.
Details can be found in Ref.~\cite{SuEiSt14_150403_}.

In this approach, termed {\it hierarchy of pure states} (HOPS) \cite{SuEiSt14_150403_}, one  introduces a new operator for the whole memory integral in Eq.~(\ref{EvolEqCorrelatorWithNoise}) and defines
\begin{equation}
\label{eq:G_t_HOPS}
 \mathcal{G}^{(e_{nj})}(t,\boldsymbol{z}^*) \equiv \Bigg( \int_{-\infty}^{\infty}ds\,\alpha_{nj}(t-s)\frac{\delta}{\delta z_{s,n}^*} \Bigg) G(t,\boldsymbol{z}^*).
\end{equation}
The integral boundaries $0$ and $t$ from Eq.~(\ref{EvolEqReducedPropagator}) are recovered later using the initial condition for $G(t,\boldsymbol{z}^*)$ and requiring causality.
The label $(e_{nj})$ in Eq.~(\ref{eq:G_t_HOPS}) is a matrix that has dimension (number of monomers) $\times$ (number of terms in the bath correlation function). It has a $1$ at position $(n,j)$ and is zero otherwise.
In the following we will map this matrix onto a vector which we denote by $\vec{e}_{nj}$.
For these operators $\mathcal{G}^{(\vec{e}_{nj})}(t,\boldsymbol{z}^*)$ one can again derive an evolution equation that leads to a (formally exact) hierarchy of coupled equations of motion.

For a practical implementation we expand/approximate the bath correlation functions $\alpha_n(t-s)$ given in Eq.~(\ref{eq:def:alpha_with_lambda}) as sums of exponentials
\begin{equation}
\label{eq:def:alpha_decomposition}
 \alpha_n(t-s) = \sum_j \alpha_{nj}(t-s),
\end{equation}
where
\begin{equation}
\label{eq:def:alpha_nj}
\alpha_{nj}(\tau) \equiv p_{nj} e^{i\omega_{nj}\tau}; \quad \tau \ge 0 
\end{equation}
with complex frequencies $\omega_{nj}= \Omega_{nj}+ i \gamma_{nj}$  and  prefactors $p_{nj}$ that may also be complex.
As demonstrated in Ref.~\cite{RiEi14_094101_}, such a decomposition can be made to handle ohmic as well as superohmic spectral densities at finite temperatures by using appropriate fitting routines.

For the bath correlation functions defined in Eqs.~(\ref{eq:def:alpha_decomposition}) and~(\ref{eq:def:alpha_nj}) one finds
\begin{align}
\label{eq:HOPS_Exps}
  & \partial_t\mathcal{G}^{(\kk)}(t,\boldsymbol{z}^*) = \nonumber \\
  & \quad \Big( -i\op{H_{\rm sys}} + i \sum_{nj} k_{nj}\omega_{nj} + \sum_n \op{K}_n\ZZ_{t,n} \Big)\mathcal{G}^{(\kk)}(t,\boldsymbol{z}^*) \nonumber \\
  & \quad + \sum_n \op{K}_n\sum_j k_{nj} \alpha_{nj}(0) \mathcal{G}^{(\kk-\ee_{nj})}(t,\boldsymbol{z}^*) \\
  & \quad - \sum_n \op{K}_n \sum_j\mathcal{G}^{(\kk+\ee_{nj})}(t,\boldsymbol{z}^*). \nonumber
\end{align}
The vector $\kk$ in Eq.~(\ref{eq:HOPS_Exps}) labels the different orders of the hierarchy (we speak of a term of $k$th order when $\sum_{n,j} k_{nj}=k$).
In Eq.~(\ref{eq:HOPS_Exps}) the $k$th order terms are connected to terms that differ by one order (as explained above $\ee_{nj}$ denotes the ($nj$)th unit vector, where $n$ labels the monomers and $j$ the terms of the bath correlation function).
This scheme (applied to wave functions instead of the reduced propagators) has been developed in \cite{SuEiSt14_150403_}.
There, also an appropriate closure of the hierarchy is discussed.

Since all the $\mathcal{G}^{(\kk)}(t,\boldsymbol{z}^*)$ are analytic functions in $z_t^*$, one obtains (see also the argument leading from Eq.~(\ref{eq:_func_expansion_o-op}) to Eq.~(\ref{EvolEqCorrelatorWithoutNoise})) an equation for $\mathcal{C}(t)$ which is independent of the time-dependent complex numbers (stochastic processes) $z_{t,n}^*$.
With the definition
 \begin{equation}
\mathcal{C}^{(\kk)}(t)=\int \frac{d^2 \boldsymbol{z}}{\pi}\,e^{-|\boldsymbol{z}|^2} \mathcal{G}^{(\kk)}(t,\boldsymbol{z}^*)
\end{equation}
one finds
\begin{equation}
\label{eq:C(t)_HOPS}
\begin{split}
  \partial_t\mathcal{C}^{(\kk)}(t) =
  &   \big( -i\op{H_{\rm sys}} +i \sum_{nj} k_{nj} \omega_{nj} \big)\,\mathcal{C}^{(\kk)}(t) \\
  & + \sum_n \op{K}_n \sum_j k_{nj} \alpha_{nj}(0) \mathcal{C}^{(\kk-\ee_{nj})}(t) \\
  & - \sum_n \op{K}_n \sum_j \mathcal{C}^{(\kk+\ee_{nj})}(t),
\end{split}
\end{equation}
and the desired correlation operator $\mathcal{C}(t)$ is 
\begin{equation}
\mathcal{C}(t)=\mathcal{C}^{(\vec{0})}(t).
\end{equation}

Note that from Eq.~(\ref{eq:C(t)_HOPS}) it is apparent that for Hermitian coupling operators $K_n$ the (finite temperature) bath correlation functions $\alpha_n(\tau)$ given by Eq.~(\ref{eq:def:alpha_with_lambda}) are the only ``environmental quantities'' that are needed for the open system dynamics. In Eq.~(\ref{eq:def:alpha_with_lambda}) only the original positive frequency oscillators enter, so that in practice the doubling of the Hilbert space for the treatment of finite temperatures does not lead to increased numerical difficulties or a worse scaling of the method.
However, for non-Hermitian coupling operators $K_n$ one truly needs two separate bath correlation functions per monomer as explained in Appendix~\ref{sec:general_equations_of_motion}.

\subsubsection{Propagation of vectors instead of a matrix}
\label{sub:vector_propagation}
Note that one can also obtain $\mathcal{C}(t)$ by independently propagating the initial states $\ket{\pi_m}$.  The matrix elements $\mathcal{C}_{nm}$ are then given by $\braket{\pi_n}{\psi_{m}(t)}$, where $ \ket{\psi_{m}(t)}$  denotes a vector obtained by propagating Eq.~(\ref{eq:C(t)_HOPS}) or Eq.~(\ref{EvolEqCorrelatorWithoutNoise}) for a \emph{vector} instead of a matrix and with the initial condition $\ket{\psi_{m}(0)}=\ket{\pi_m}$.
(See also Appendix~\ref{sec:Dimer}, where we use this scheme explicitly for a molecular dimer.)

\section{Example calculations}
\label{sec:example_calculations}

In this section we present calculated absorption and CD spectra to illustrate a typical situation for which the present approach can be used.
We do not aim at a detailed interpretation of the spectra
and also do not intend to give an investigation of the speed of convergence of HOPS with the order of the hierarchy, here.
General convergence properties of HOPS have been discussed in the Supplemental Material of Ref.~\cite{SuEiSt14_150403_}.

All spectra shown in this work are calculated using the HOPS approach with order $k = 10$ and can be considered to be converged \footnote{The spectra are converged in the sense that by increasing the order of the hierarchy from 9 to 10 the presented plots do not change noticeably on the presented scale. In fact already at smaller order most spectra can be considered to be converged.}.

\begin{figure}
\includegraphics[width=1.0\columnwidth]{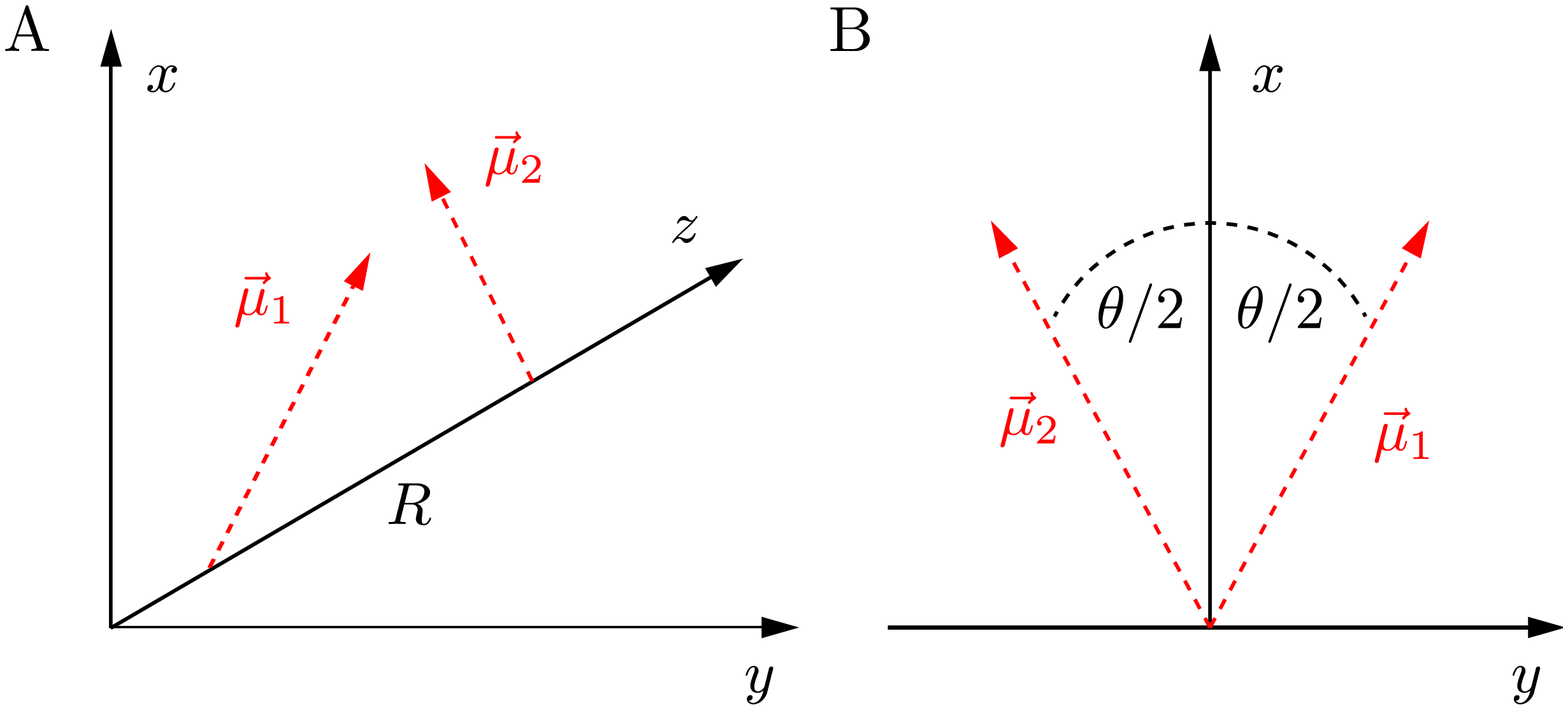}
\caption{\label{fig:dimer_sketch}
Sketch of the considered geometry.\\
A: The chosen coordinate system. The centers of the two molecules are located on the $z$-axis separated by the distance $R$. The two transition dipoles are also located at these points and perpendicular to the $z$-axis. Their directions contain all information about the molecular orientations necessary for the calculation of optical spectra.\\
B: Specification of the geometry used for the spectra shown.
}
\end{figure}

We consider the simplest case of an aggregate, i.e.\ a dimer composed of two monomers.
The interaction $V_{12} =V_{21}$ between the two monomers in Eq.~(\ref{H_sys}) we denote by $V$.
We take the electronic transition energies and spectral densities to be equal for both monomers ($\varepsilon_n=\varepsilon$ and $\alpha_n(\tau)=\alpha(\tau)$) and consider their local environments to be uncorrelated.
Since in this work our concern is not the detailed investigation of the dependence of the spectra on the orientation of the monomers, we restrict the geometry of the dimer by taking the transition dipoles of the monomers perpendicular to the distance vector $\vec{R}\equiv\vec{R}_{12}$ between them.
Then the angle $\theta$ between the monomers is sufficient to describe the geometry, which is sketched in Fig.~\ref{fig:dimer_sketch}.
If $\theta$ is not equal to $0$ or $\pi$ then the system has chirality and shows a CD signal.

\begin{figure}
\includegraphics[width=\columnwidth]{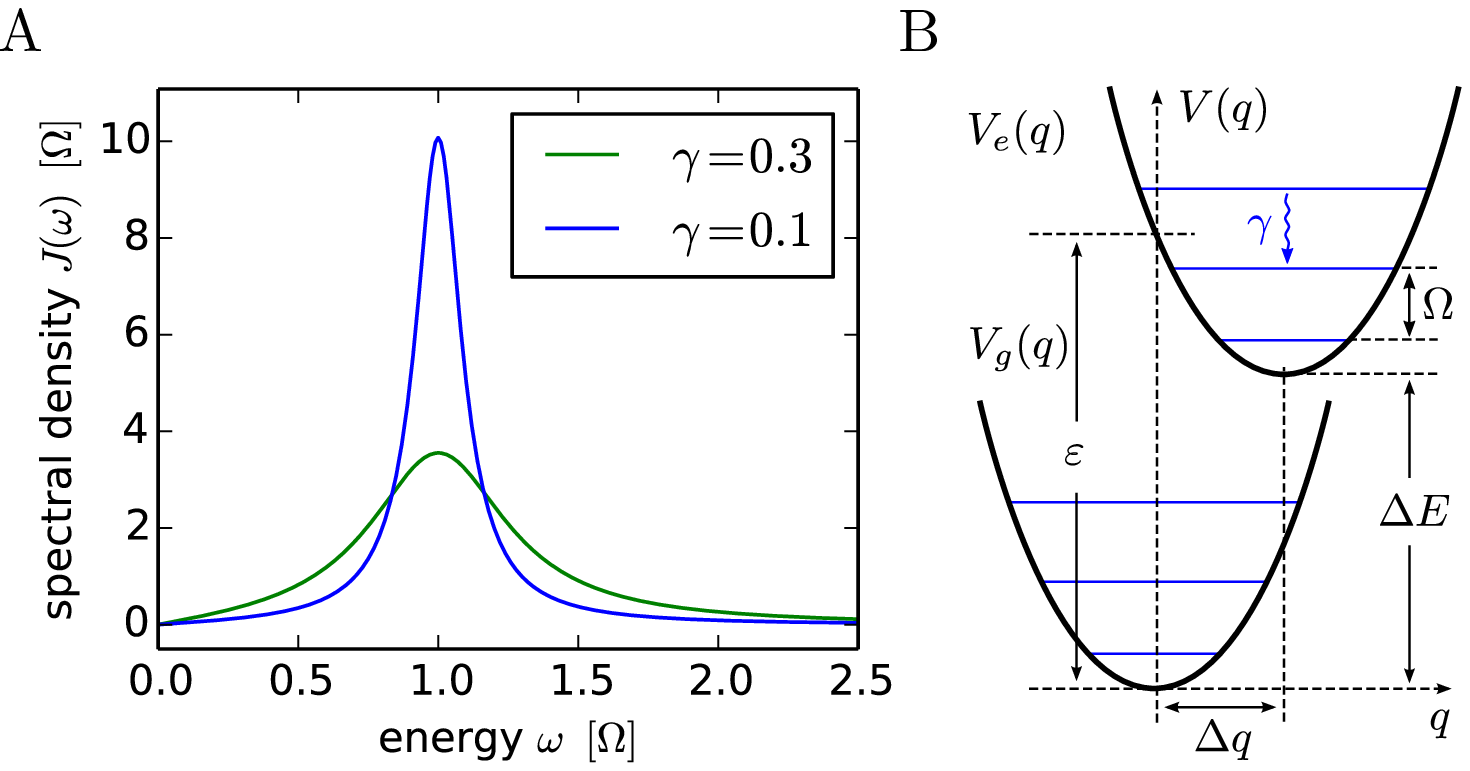}
\caption{
A: Spectral densities $J(\omega)$ for the reorganization energy $E_r=1.0$ and two different width parameters $\gamma$ (all energies are in units of $\Omega$).\\
B: Possible interpretation in terms of shifted harmonic Born-Oppenheimer surfaces for one vibrational coordinate $q = \frac{1}{\sqrt{2\Omega}}(a^{\dagger} + a)$. See Appendix~\ref{app:aggregate} for details.
}
\label{fig:spd}
\label{fig:shifted_HO}
\end{figure}

The coupling to the (vibrational) environment we describe by a continuous spectral density. For the example calculations shown below the spectral density is chosen to be of the simple anti-symmetrized Lorentzian form
\begin{equation}
 J(\omega) = p \Big( \frac{1}{(\omega-\Omega)^2  +\gamma^2} - \frac{1}{(-\omega-\Omega)^2  +\gamma^2} \Big),
\label{eq:_lorentzian_sd}
\end{equation}
for which the corresponding bath correlation function is known analytically \cite{MeTa99_3365_} as a sum of exponentials of the form Eq.~(\ref{eq:def:alpha_decomposition}).
 In order to calculate the bath correlation function for different temperatures, we use the method described in Ref.~\cite{RiEi14_094101_}, which is based on a Padé approximation \cite{HuXu10_101106_} of the hyperbolic cotangent appearing in Eq.~(\ref{eq:def:alpha_with_lambda}).
Note that in Ref.~\cite{RiEi14_094101_} a much larger class of spectral densities was suggested that allow for an analytic determination of the bath correlation function as such a sum of exponentials. Especially, the spectral densities considered there allow one to describe superohmic behavior near $\omega=0$.
However, in this work we restrict ourselves to the case Eq.~(\ref{eq:_lorentzian_sd}). This spectral density Eq.~(\ref{eq:_lorentzian_sd}) shows a linear increase for small $\omega$ and falls off with the third power for large $\omega$.
It can be approximately interpreted as describing the coupling of an electronic transition to a damped molecular vibration with frequency $\Omega$, where the damping is described by the constant $\gamma$.
Problems of this interpretation are discussed in Ref.~\cite{RoStWh12_204110_}.
The prefactor $p$ is connected to the so-called reorganization energy $E_r=\frac{1}{\pi} \int_0^{\infty} d\omega\, \frac{J(\omega)}{\omega}$
 by
$p=E_r \gamma (\Omega^2+\gamma^2)/\Omega $.
Fig.~\ref{fig:spd} shows a plot of the spectral density for two different width parameters $\gamma$.
All quantities are expressed in units of the central frequency $\Omega$.

In the following we present spectra of dimers as well as the corresponding spectra of the uncoupled monomers.
Note that the parameters in this work have been deliberately chosen such that the previously used ZOFE approximation (see Appendix~\ref{sec:ZOFE-approach}) does not give reliable results in most of the cases (i.e.\ we choose a relatively strong reorganization energy $E_r$ and a small environmental damping $\gamma$).

\subsection{Temperature-dependent spectra}
For linear optical properties of molecular aggregates, many general results for the interpretation of absorption and CD can be found in the book by Rodger and Nord\'{e}n \cite{RoNo97__}.
For the special case of a dimer with internal vibrations of the monomers we refer in particular to Ref.~\cite{EiSeEn08_186_}.

\begin{figure}[tp]
\includegraphics[width=1.0\columnwidth]
{{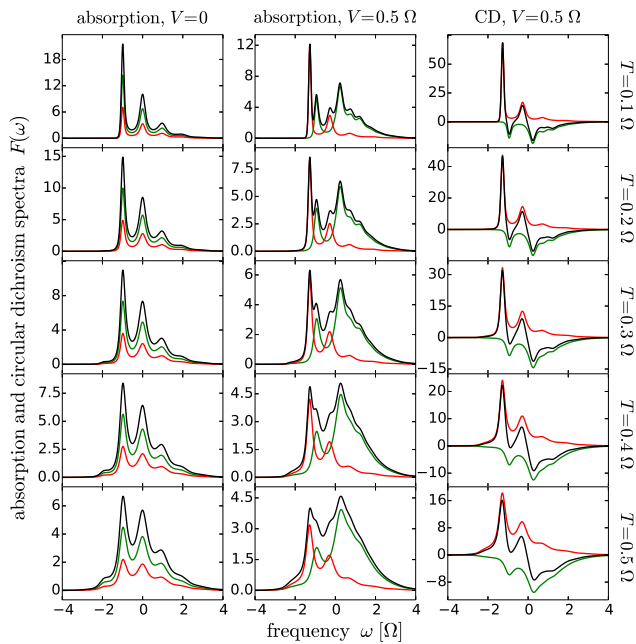}}
\caption{\label{fig:temp_dep_01}
Temperature dependence of absorption for monomer (left column) and dimer (middle column) as well as CD (right column). Parameters are $E_r=1\ \Omega$, $\gamma=0.1\ \Omega$, and $V=0.5\ \Omega$ (for the dimer).
The temperatures are from top to bottom: $T/\Omega = $ $0.1, 0.2, 0.3, 0.4, 0.5$.
The red and green curves show the two weighted contributions from the symmetric and antisymmetric parts of the dimer wave function (see Appendix~\ref{sec:Dimer}) and the black curves show the resulting absorption and CD spectra.
The zero on the abscissa is chosen as the center of mass of the monomer spectrum in each row.
Absorption spectra are given in units of $\mu^2$ and CD spectra in units of $R\,\mu^2$.
}
\end{figure}

\begin{figure}[tp]
\includegraphics[width=1.0\columnwidth]
{{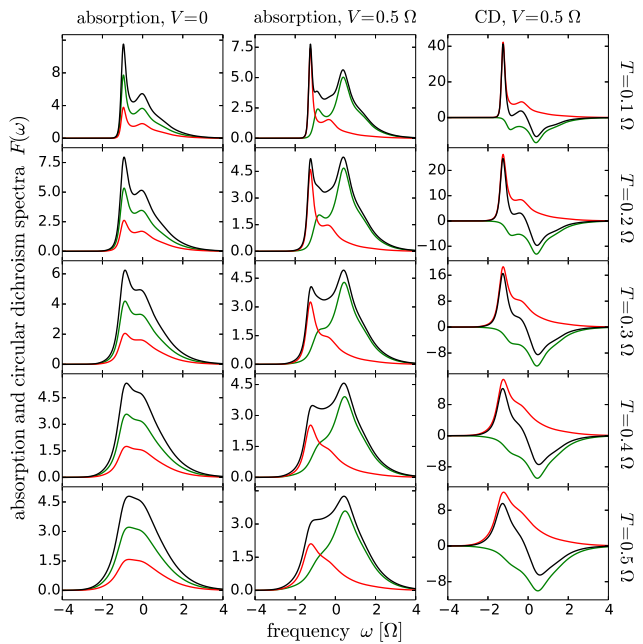}}
\caption{\label{fig:temp_dep_03}
Same as Fig.~\ref{fig:temp_dep_01}, but for $\gamma=0.3$.}
\end{figure}

To demonstrate the influence of temperature we present some exemplary spectra that we obtain for the spectral density Eq.~(\ref{eq:_lorentzian_sd}) at various temperatures.
We take two Padé expansion terms for the hyperbolic cotangent into account for all temperatures. Only for $T=0.1$ and $\gamma=0.3$ we take three expansion terms.
We show absorption and CD spectra for two uncoupled (monomer case) and two coupled (dimer case) monomers at different temperatures for two spectral densities with width parameter $\gamma=0.1$ (Fig.~\ref{fig:temp_dep_01}) and  $\gamma=0.3$  (Fig.~\ref{fig:temp_dep_03}).
In the left column of each figure the linear absorption for $V=0$, i.e.\ for uncoupled monomers, is shown. In the middle and right columns we plotted linear absorption and CD spectra for a dimer with $V=0.5$ and $\theta=70^{\circ}$.
For non-interacting monomers, which do not possess a CD of their own, the CD vanishes, as expected.
It is well known (and for the present formalism stated in Appendix~\ref{sec:Dimer}) that for the chosen geometry the absorption and CD spectrum can be understood by considering the correlation functions of the symmetric and antisymmetric parts of the dimer wave function.
In the absorption spectrum the angle $\theta$ enters as a weighting of the two contributions, in the CD spectra $\theta$ only scales the absolute values.
Note that if the interaction between the monomers is of point-dipole-dipole type, then $\theta$ also enters in the interaction strength via $V \sim (1-3 \cos^2\theta)/R^3$.
We have chosen the sign of $V$ to be consistent with this type of interaction.
In the plots we show the symmetric and anti-symmetric contributions to the spectra (weighted by their $\theta$-dependent prefactor), which we obtain from Eq.~(\ref{eq:_c_abs_plus_minus}) for absorption and Eq.~(\ref{eq:_c_cd_plus_minus}) for CD, as colored lines (red and green, respectively) and the resulting absorption and CD spectra as black curves.
The absorption spectra are normalized to have an area of $2\pi$.
Note that the integral over the CD spectra vanishes, as it should.
In absorption, the red curve can be understood as the absorption of light polarized along the $x$-axis and the green curve as the spectrum for light polarized along the $y$-axis.

First we take a brief look at the situation $V=0$, i.e.\ uncoupled monomers, depicted in the left columns of Figs.~\ref{fig:temp_dep_01} and~\ref{fig:temp_dep_03}.
In Fig.~\ref{fig:temp_dep_01} we see for the monomer case a `vibrational progression' corresponding to the (broadened) mode $\Omega$ of the spectral density (peaks at $\omega =-1,0,1,2$). With increasing temperature (from top to bottom) the spectrum becomes broader. Additionally, absorption at lower energies than the one of the dominating 0-0 transition of the mode $\Omega$ (peak at $\omega=-1$) increases (hot bands).
The black, red and green curve obviously have the same shape.

Let us now consider the dimer case.
We choose a positive dipole-dipole coupling $V=0.5$ and an angle $\theta=70^{\circ}$ between the monomer transition dipoles. Absorption and CD spectra for this case are shown in the middle (absorption) and right (CD) columns of Figs.~\ref{fig:temp_dep_01} and~\ref{fig:temp_dep_03}.
One sees that the dimer spectra contain a lot more structure compared with the monomer spectra.
This complicates the interpretation of measured spectra, since one might for example be tempted to interpret the second peak from the left in the dimer absorption spectra as belonging to a vibrational progression with the lowest peak.
Methods to extract the usually unknown geometry of the dimer with many vibrational modes from absorption and CD are discussed in Refs.~\cite{Ei07_321_, EiSeEn08_186_}.
As in the monomer case increasing the temperature leads to a broadening of the spectra.

For the larger value of the width parameter $\gamma$ we see in Fig.~\ref{fig:temp_dep_03} that the features in the spectra are broadened and smeared out already for the lower temperatures so that less structure of the spectra is visible, as one expects.

\section{Conclusions}
\label{sec:conclusions}
In this work we have shown how to calculate linear optical spectra (in particular absorption and CD) of molecular aggregates using the NMQSD formalism.
One key aspect was to demonstrate that, while the NMQSD formalism in general requires to calculate many stochastic trajectories, for the calculation of linear optical properties one deterministic propagation is sufficient.
To this end we have mapped the finite temperature case to effective temperature-zero equations for which the stochastics drops out.

We  summarize the relevant equations that we used to calculate the transition strength $F(\omega)$ for linear absorption and CD spectra.
The transition strengths are calculated from Eq.~(\ref{eq:spectrum}) with the correlation function Eq.~(\ref{eq:def:C_abs}) for absorption and Eq.~(\ref{eq:def:C_CD}) for CD. 
These formulas are for isotropically oriented samples. However, one can also easily treat the case of oriented samples and light of various polarizations~(see Eq.~(\ref{eq_app::c_t_full}) in Appendix~\ref{sec:c(t)}).
Similar to the treatment in Ref.~\cite{EiKnBr07_104904_} one can also go beyond the long wavelength approximation.
In all cases one has to calculate the same correlation operator $\mathcal{C}(t)$ given in its general form in Eq.~(\ref{EvolEqCorrelatorWithNoise2}).
To solve this equation, in the present work we used the HOPS formalism, which leads to
the hierarchy of correlation operators Eq.~(\ref{eq:C(t)_HOPS}) whose solution is equivalent to the solution of Eq.~(\ref{EvolEqCorrelatorWithNoise2}).
Here, one sees explicitly that a single trajectory with $z^*_{t,n}\equiv 0$ is sufficient to obtain the required spectra.

In a similar manner one can show that also for the ZOFE approximation (for details see Appendix~\ref{sec:ZOFE-approach}) only a single trajectory is required.
While HOPS allows for a numerically exact treatment, ZOFE is computationally much less demanding and yields in many situations of interest very good agreement with the exact result.

As application of the presented formulas we have in mind aggregates typically composed of organic molecules (monomers).
In each monomer the electronic excitation naturally couples to nuclear vibrations of the molecule. Typically, due to interaction with external vibrations, those intramolecular vibrations are damped, which manifests itself in broadened peaks in the spectral density. This then leads in general to broadened vibrational progressions in the absorption spectra of the single molecules.
In molecular aggregates, however, the electronic interaction between the molecules leads to non-trivial spectra, where the identification of vibrational progressions can be complicated.

The method presented in this work is well-suited to calculate optical properties of molecular aggregates at finite temperatures with explicit inclusion of vibrational modes.

We have shown exemplarily for the case of a dimer how our method can be used.


\appendix


\section{The aggregate Hamiltonian}
\label{app:aggregate}

In this appendix we give a very brief derivation of the model we used.
For a more complete overview, we refer, e.g., to the book by May and K\"uhn \cite{MaKue11__}.

\subsection{The monomer}
\label{sec:monomer}

For each monomer $n$ we take into account its electronic ground state $\ket{\phi_n^g}$ 
and one excited electronic state $\ket{\phi_n^e}$.
The transition energy between these two states is denoted by $\varepsilon_n$.
 Each monomer has a collection of vibrational modes 
comprising internal ones as well as modes of the local environment of the chromophore.
We will refer to these degrees of freedom  as ``the bath''.
We choose the  vibrational modes to be harmonic and consider a linear coupling of the electronic excitation of monomer $n$ to these bath degrees of freedom (making contact to 
previous work ~\cite{WiMo60_872_,Me63_154_,FuGo64_2280_,Wi65_161_,ScFi84_269_,EiBrSt05_134103_,SeMaEn06_354_,WaEiBr08_044505_}).
The Hamiltonian of monomer $n$ is then given by
\begin{equation}
\label{app:H_n}
  H_n=H^g_n\ket{\phi_n^g}\bra{\phi_n^g}+H^e_n\ket{\phi_n^e}\bra{\phi_n^e},
\end{equation} 
with the Hamiltonian of the vibrations in the electronic ground state
\begin{equation}
\label{HamMonGround}
  H^g_n=\sum_{\lambda}\omega_{n\lambda}a^{\dagger}_{n\lambda}a_{n\lambda}
\end{equation}
(the energies of the vibrational ground states of all modes in the electronic ground 
state are chosen to be zero). The excited state Hamiltonian of monomer $n$ is written as
\begin{equation}
\label{HamMonExc}
  H^e_n = \varepsilon_n + \sum_{\lambda}\omega_{n\lambda}a^{\dagger}_{n\lambda}a_{n\lambda} - \sum_{\lambda}\kappa_{n\lambda}(a^{\dagger}_{n\lambda}+a_{n\lambda}).
\end{equation}
Here $a_{n\lambda}^{\dagger}$ ($a_{n\lambda}$) denotes the creation (annihilation) operator of the local mode $\lambda$ of monomer $n$ with frequency $\omega_{n\lambda}$.
The corresponding coupling strength between the electronic excitation of monomer $n$ and this mode is denoted by $\kappa_{n\lambda}$.
The resulting Hamiltonian $H_n$ can be interpreted in terms of two multidimensional shifted harmonic potential energy surfaces (see Fig.~\ref{fig:shifted_HO}). This interpretation is discussed, e.g., in Refs.~\cite{MaKue11__, RoStWh12_204110_}.

If the spectral density consists of only a few modes $\lambda$ (or equivalently, it is a sum of a few delta-peaks), these modes can be considered as coming from a model of the monomer that consists of (multitimensional) harmonic Born-Oppenheimer surfaces where ground and excited state have the same shape but are shifted with respect to each other.
The freqencies appearing in the spectral density are then simply the frequencies of the harmonic potentials (in the different directions) and the coupling strength $|\kappa_{n\lambda}|^2$ is related to the shift of the potentials \cite{MaKue11__}.

Peaks in the spectral density that are not delta-like but broadened can arise due to internal vibrational modes (as described above) that are damped by an environment. 
This model is discussed, e.g., in Ref.~\cite{RoStWh12_204110_}.

Beside internal vibrations of the monomers also environmental fluctuations can lead to peaks in the spectral density.
This is discussed, e.g., in Refs.~\cite{Mu95__,DaKoKl02_031919_,VaEiAs12_224103_}.

\subsection{The aggregate}

The total Hamiltonian of the aggregate is a sum of the Hamiltonians Eq.~(\ref{app:H_n})  of the individual monomers and the dipole-dipole interaction operator between them.
Taking matrix elements in the basis Eq.~(\ref{ElGroundStateAgg}) and Eq.~(\ref{pi_n_states}) we find for the electronic ground state Hamiltonian 
\begin{equation}
   H_{\bath}=\sum_{n=1}^N H_n^g =\sum_{n=1}^N\sum_{\lambda}\omega_{n\lambda}a^{\dagger}_{n\lambda}a_{n\lambda} 
\end{equation}
and for the Hamiltonian in the electronically excited state
\begin{equation}
\label{HamTotExc}
\begin{split}
   H^e=\sum_{n=1}^N &\Big(H_n^e+\sum_{\ell\ne n}^N H_{\ell}^g\Big)\ket{\pi_n}\bra{\pi_n}+ \sum_{n,m=1}^N V_{nm}\ket{\pi_n}\bra{\pi_m}.
\end{split}
\end{equation}
with the matrix elements $V_{nm}$ of the interaction operator which are taken to be independent of environmental (nuclear) coordinates.

Note that $H_n^g$ and $H_n^e$ depend on nuclear coordinates through Eqs.~(\ref{HamMonGround}) and (\ref{HamMonExc}).
With the Hamiltonians of the monomers Eqs.~(\ref{HamMonGround}) and (\ref{HamMonExc}) we can write Eq.~(\ref{HamTotExc}) as Eq.~(\ref{HeAsSysPlusIntPlusEnv}) of the main text.

\section{Correlation function}
\label{sec:c(t)}

In this section, the expressions Eqs.~(\ref{eq:def:C_abs})~and~(\ref{eq:def:C_CD}) for the correlation function entering Eq.~(\ref{eq:spectrum}) for the absorption strength are derived.
For linear absorption the orientation averaged dipole-autocorrelation function that needs to be evaluated \cite{Mu95__} reads  
\begin{equation}
\label{eq_app:c_t_muk}
 c^{\rm Abs}(t) = \tr \{ \hat\mu(t)\cdot\hat\mu(0)\, \rho_0 \},
\end{equation}
with $\hat\mu(t)$ being the corresponding Heisenberg operator for the dipole operator of the aggregate defined in Eq.~(\ref{eq:dip_op_agg}) and $\rho_0$ the total initial density operator of system and environment defined in Eq.~(\ref{eq:rho_0}). The dot denotes the scalar product of the vector operators in Eq.~(\ref{eq_app:c_t_muk}), where the time-dependence of the Heisenberg operator $\hat\mu(t)$  is given by
\begin{equation}
\hat\mu(t) = U^\dagger(t)\, \hat\mu\,  U(t)
\end{equation}
with $U(t) = e^{-i H t}$ and $H$ being the full aggregate Hamiltonian defined in Eq.~(\ref{eq:H_agg}).
The matrix elements of the (time-independent) dipole operator $\hat\mu$ between the basis states Eqs.~(\ref{ElGroundStateAgg})\ and\ (\ref{pi_n_states}) are
\begin{equation}
\label{eq:matrix_elements_mu}
 \bra{g_{\rm el}}\hat\mu\ket{\pi_n} = \vec\mu_n \quad\textrm{and}\quad\bra{\pi_n}\hat\mu\ket{\pi_m} = 0.
\end{equation}
Inserting Eq.~(\ref{eq:rho_0}) for $\rho_0$ into Eq.~(\ref{eq_app:c_t_muk}) and evaluating the partial trace over the electronic system degrees of freedom yields
\begin{equation}
\label{eq_app:c_t}
 c^{\rm Abs}(t) = \tr_{\rm env} \{ \bra{g_{\rm el}}\hat\mu(t)\cdot\hat\mu(0)\ket{g_{\rm el}}\, \rho_{\rm env} \}.
\end{equation}
After expanding $\hat\mu$ and $U(t)$ in terms of the basis states $\ket{g_{\rm el}}$ and $\ket{\pi_n}$ using Eqs.~(\ref{eq:matrix_elements_mu})~and~(\ref{eq:def:U}), equation \eqref{eq_app:c_t} simplifies to
\begin{equation}
\label{eq_app::c_t_exp}
 c^{\rm Abs}(t) = \sum_{n,m} \vec\mu_n\cdot\vec\mu_m\, \mathcal{C}_{nm}(t)
\end{equation}
with $\mathcal{C}_{nm}$ given by Eq.~(\ref{eq:def:C_nm}). This is Eq.~(\ref{eq:def:C_abs}) from the main text.

For calculating CD one needs to include the polarization and relative phases of the electric field 
\begin{equation}
 \frac{\vec E(t,\vec{R})}{|\vec E(t,\vec R)|} = \vec{\varepsilon}\; e^{-i {\vec k\cdot \vec R}} e^{i\omega t} 
\end{equation}
with $\vec \varepsilon$ the complex valued polarization vector ($|\vec \varepsilon| = 1$) and $\vec k$ the wave vector.
Following the above derivation the final expression now reads
\begin{equation}
\label{eq_app::c_t_full}
 c_{\vec{\varepsilon},\vec{k}}^{\rm Abs}(t) = \sum_{n,m} (\vec\mu_n \cdot {\vec{\varepsilon}}\,\raisebox{0.9ex}{\footnotesize *} e^{i {\vec k\cdot \vec R_m}})(\vec\mu_m \cdot \vec{\varepsilon}\, e^{-i {\vec k\cdot \vec R_n}})\, \mathcal{C}_{nm}(t).
\end{equation}
The CD signal is obtained by taking the difference of left and right polarized light.

In the often used Rosenfeld formalism \cite{Ro29_161_}, which essentially corresponds to the approximation $e^{i \vec{k}\cdot\vec{R}}\approx 1+i \vec{k}\cdot\vec{R}$ the expression for CD is 
\begin{equation}
\label{eq_app::c_t_CD}
 c^{\rm CD}(t) = \sum_{nm} \vec{R}_{nm}\cdot (\vec\mu_n \times \vec\mu_m)\, \mathcal{C}_{nm}(t),
\end{equation}
with $\vec{R}_{nm} = \vec R_m - \vec R_n$ and $\mathcal{C}_{nm}(t)$ as defined above. Here again an orientational averaging was done to get the final expression \eqref{eq_app::c_t_CD}, which is Eq.~(\ref{eq:def:C_CD}) from the main text.

Note that applying the long wavelength approximation and performing an orientational averaging of \eqref{eq_app::c_t_full} yields again Eq.~\eqref{eq_app::c_t_exp}.

\section{Equations of motion for the propagator}
\label{sec:general_equations_of_motion}

We insert the completeness relations Eq.~(\ref{eq:completeness_Bargmann}) in Eq.~(\ref{eq:CorrelationOperatorThermofield}) for the correlation operator and obtain
\begin{equation}
\label{eq:correlation_operator_with_G_of_zeta_xi}
\begin{split}
 \mathcal{C}(t) &= \int \frac{d^2 \boldsymbol{\zeta}}{\pi}\,e^{-|\boldsymbol{\zeta}|^2}\, \int \frac{d^2 \boldsymbol{\xi}}{\pi}\,e^{-|\boldsymbol{\xi}|^2}\, G(t,\boldsymbol{\zeta}^*, \boldsymbol{\xi}^*),
\end{split}
\end{equation}
where we defined the reduced propagator
\begin{equation}
\label{eq:def:G_zeta_xi}
 G(t,\boldsymbol{\zeta}^*, \boldsymbol{\xi}^*) \equiv \bra{\boldsymbol{\zeta}}_A \bra{\boldsymbol{\xi}}_B \bar{U}^{e}(t) \ket{0(\beta)}.
\end{equation}
To obtain Eq.~(\ref{eq:def:G_zeta_xi}), we have made use of $\bra{0(\beta)} \big( \ket{\boldsymbol{\zeta}}_A \ket{\boldsymbol{\xi}}_B \big) = 1$.
$G(t,\boldsymbol{\zeta}^*, \boldsymbol{\xi}^*)$ is analytic in $\boldsymbol{\zeta}^*$ and $\boldsymbol{\xi}^*$ because of the properties of the Bargmann states \cite{Ba61_187_}.
Therefore one can Taylor expand it with respect to these parameters resulting in
\begin{equation}
\label{eq:G_zeta_xi_Taylor}
G(t,\boldsymbol{\zeta}^*, \boldsymbol{\xi}^*)= G^{(0)}(t)
+\sum_{n,\lambda} \big(G^{(1_\xi)}_{n\lambda}(t)\xi_{n\lambda}^{*}+G^{(1_\zeta)}_{n\lambda}(t)\zeta_{n\lambda}^{*}) +\dots
\end{equation}
Inserting this into Eq.~(\ref{eq:correlation_operator_with_G_of_zeta_xi}) and noting that  $\int\frac{d^2 \xi_{n\lambda}}{\pi}\,e^{-|\xi_{n\lambda}|^2} (\xi^{*}_{n\lambda})^j = \delta_{j0}$ (the same also holds for integrals containing $\zeta^{*}_{n\lambda}$), one sees that only the zeroth-order term $G^{(0)}(t)$ survives and one obtains
\begin{equation}
\label{eq:C(t)=G^{(0)}(t)}
 \mathcal{C}(t)=G^{(0)}(t).
\end{equation}
It has been shown in Ref.~\cite{DiGiSt98_1699_} that there exists an integro-differential equation for $G(t,\boldsymbol{\zeta}^*, \boldsymbol{\xi}^*)$ which can be used as a starting point to obtain $G^{(0)}(t)$.

We will now first present an evolution equation for the general propagator $G(t,\boldsymbol{\zeta}^*, \boldsymbol{\xi}^*)$ defined in Eq.~(\ref{eq:def:G_zeta_xi}).
Since that equation holds even for system coupling operators that are not necessarily Hermitian, we use the symbol $L_n$ in Appendix~\ref{sub:general_propagator} in order to distinguish them from the Hermitian operators $K_n$.
After presenting the general equations, we treat the special case of Hermitian system-bath coupling operators $K_n = K_n^{\dagger}$ and derive Eq.~(\ref{EvolEqReducedPropagator}).

\subsection{General propagator}
\label{sub:general_propagator}
We start by differentiating Eq.~(\ref{eq:def:G_zeta_xi}) with respect to time and obtain the evolution equation
\begin{equation}
\label{eq:EvolEqGWithU}
 \partial_t G(t,\boldsymbol{\zeta}^*, \boldsymbol{\xi}^*) = \bra{\boldsymbol{\zeta}}_A \bra{\boldsymbol{\xi}}_B \big( -i\bar{H}^{e} \bar{U}^{e}(t) \big) \ket{0(\beta)}
\end{equation}
with $\bar{H}^e$ given in Eq.~(\ref{eq:bar_H_e}).
(Note: Unlike the full excited state propagator $\bar{U}^{e}(t)$ the reduced propagator $G(t,\boldsymbol{\zeta},\boldsymbol{\xi})$ obtained by projecting on the environmental degrees of freedom is no longer unitary.)

After a transformation to the interaction picture and using the properties of the Bargmann representation one obtains the equation of motion
\begin{equation}
\label{eq:EvolEqReducedPropagator_lambda}
\begin{split}
\partial_t G(t,\boldsymbol{\zeta}^*, \boldsymbol{\xi}^*) =
& -i H_{\rm sys}G(t,\boldsymbol{\zeta}^*, \boldsymbol{\xi}^*) \\
&\hspace{-1.5cm} -i\sum_n L_n\sum_{\lambda}  (g_{n\lambda}^+)^* e^{i\omega^+_{n\lambda} t}\zeta_{n\lambda}^*G(t,\boldsymbol{\zeta}^*, \boldsymbol{\xi}^*)\\
&\hspace{-1.5cm}  -i\sum_n L^{\dagger}_n \sum_{\lambda} (g_{n\lambda}^-)^*  e^{i\omega^-_{n\lambda} t} \xi^*_{n\lambda}G(t,\boldsymbol{\zeta}^*, \boldsymbol{\xi}^*) \\
&\hspace{-1.5cm} -i\sum_n L^{\dagger}_n  \sum_{\lambda} (g_{n\lambda}^+)e^{-i\omega^+_{n\lambda} t} \frac{\partial}{\partial \zeta^*_{n\lambda}} G(t,\boldsymbol{\zeta}^*, \boldsymbol{\xi}^*) \\
&\hspace{-1.5cm} -i\sum_n L_n  \sum_{\lambda} (g_{n\lambda}^-)e^{-i\omega^-_{n\lambda} t} \frac{\partial}{\partial \xi^*_{n\lambda}} G(t,\boldsymbol{\zeta}^*, \boldsymbol{\xi}^*) \\
\end{split}
\end{equation}
with the not necessarily Hermitian system coupling operators $L_n$ instead of the $K_n$ in Eq.~(\ref{HInt}).
In Eq.~(\ref{eq:EvolEqReducedPropagator_lambda}) we have defined
\begin{align}
g_{n\lambda}^+ = &\sqrt{\bar{n}_{n\lambda}+1}\ \kappa_{n\lambda},\quad\quad\quad  \omega_{n\lambda}^+=\omega_{n\lambda},\\
g_{n\lambda}^-=&\sqrt{\bar{n}_{n\lambda}}\ \kappa_{n\lambda}, \quad\quad\quad\quad\quad  \omega_{n\lambda}^-=-\omega_{n\lambda}.
\end{align}

As shown in Refs.~\cite{DiGiSt98_1699_, Yu04_062107_} one can pave the way to an efficient stochastic treatment by rewriting this equation according to 
\begin{equation}
\label{eq:EvolEqReducedPropagator}
\begin{split}
\partial_t G(t,\boldsymbol{\zeta}^*, \boldsymbol{\xi}^*) =
& -i H_{\rm sys}G(t,\boldsymbol{\zeta}^*, \boldsymbol{\xi}^*) \\
&\hspace{-1.5cm} +\sum_n L_n \zeta^*_{t,n}G(t,\boldsymbol{\zeta}^*, \boldsymbol{\xi}^*)
  +\sum_n L^{\dagger}_n \xi^*_{t,n}G(t,\boldsymbol{\zeta}^*, \boldsymbol{\xi}^*) \\
&\hspace{-1.5cm} -\sum_n L^{\dagger}_n \int_0^t ds\ \alpha_{1,n}(t-s)\frac{\delta}{\delta \zeta^*_{s,n}} G(t,\boldsymbol{\zeta}^*, \boldsymbol{\xi}^*) \\
&\hspace{-1.5cm} -\sum_n L_n \int_0^t ds\ \alpha_{2,n}(t-s)\frac{\delta}{\delta \xi^*_{s,n}} G(t,\boldsymbol{\zeta}^*, \boldsymbol{\xi}^*)
\end{split}
\end{equation}
with initial condition $G(0,\boldsymbol{\zeta}^*,\boldsymbol{\xi}^*) = \openone$. In Eq.~(\ref{eq:EvolEqReducedPropagator}) $\zeta^*_{t,n}$ and $\xi^*_{t,n}$ are time-dependent complex numbers given by
\begin{equation}
\label{eq:def:zeta_xi}
\begin{split}
 \zeta^*_{t,n} &= -i\sum_{\lambda}\sqrt{\bar{n}_{n\lambda}+1}\ \kappa_{n\lambda}\zeta^{*}_{n\lambda}e^{i\omega_{n\lambda t}}, \\
 \xi^*_{t,n} &= -i\sum_{\lambda}\sqrt{\bar{n}_{n\lambda}}\ \kappa_{n\lambda}\xi^{*}_{n\lambda}e^{-i\omega_{n\lambda t}}
\end{split}
\end{equation}
and $\alpha_{1,n}$, $\alpha_{2,n}$ are defined as
\begin{equation}
\label{eq:def:alpha1_alpha2}
\begin{split}
 \alpha_{1,n}(t-s) &= \sum_{\lambda} (\bar{n}_{n\lambda}+1) |\kappa_{n\lambda}|^2 e^{-i\omega_{n\lambda} (t-s)}, \\
 \alpha_{2,n}(t-s) &= \sum_{\lambda} (\bar{n}_{n\lambda}) |\kappa_{n\lambda}|^2 e^{i\omega_{n\lambda} (t-s)}.
\end{split}
\end{equation}

\subsection{Hermitian coupling}
\label{sub:hermitian_coupling}
For Hermitian coupling operators $L_n=L^{\dagger}_n \equiv K_n$ the above equations simplify considerably, since one can now combine the $\xi$ and $\zeta$ terms in the summation.
Introducing the convention that the mode label $\lambda$  runs over positive integers (i.e.\ 1, 2, 3, \dots) we define new quantities labeled by $\mu =\dots,\ -3,\ -2,\ -1,\ +1,\ +2,\ +3, \dots$ according to
\begin{align}
  z_{n\mu} =
  \begin{cases}
      \zeta_{n,\mu} \\
      \xi_{n,-\mu}
  \end{cases}
;
\quad
 g_{n\mu} =
  \begin{cases}
      g^+_{n,\mu} \\ 
      g^-_{n,-\mu}
  \end{cases}
;
\quad
 \tilde{\omega}_{n\mu} =
  \begin{cases}
      \omega^+_{n,\mu} \\
      \omega^-_{n,-\mu}
  \end{cases},
\end{align}
where the upper row holds for $\mu>0$ and the lower row for $\mu<0$.

Equation \ref{eq:EvolEqReducedPropagator_lambda} can then be written as
\begin{equation}
\label{eq:EvolEqReducedPropagator_lambda_z}
\begin{split}
\partial_t G(t,\boldsymbol{z}^*) =
& -i H_{\rm sys}G(t,\boldsymbol{z}^*) \\
&\hspace{-1.5cm} -i\sum_n K_n \sum_{\mu}  g_{n\mu}^* e^{i\tilde{\omega}_{n\mu} t}z_{n\mu}^*G(t,\boldsymbol{z}^*)\\
&\hspace{-1.5cm} -i\sum_n K_n  \sum_{\mu} g_{n\mu}e^{-i\tilde{\omega}_{n\mu} t} \frac{\partial}{\partial z^*_{n\mu}} G(t,\boldsymbol{z}^*).
\end{split}
\end{equation}
To obtain Eq.~(\ref{EvolEqReducedPropagator}) we follow the procedure of Ref.~\cite{DiGiSt98_1699_} and introduce 
\begin{equation}
z^*_{t,n}=- i \sum_{\mu} g_{n\mu} e^{i \tilde{\omega}_{n\mu} t} z^*_{n\mu}
\end{equation}
Note that
$z^*_{t,n} = \xi_{t, n}^*+\zeta_{t, n}^*$.
Using $\partial/\partial  z^*_{n\mu}=\int ds\, (\partial z^*_{s,n} /\partial z^*_{n\mu}) (\delta/\delta z^*_{s,n}) $ one 
obtains Eq.~(\ref{EvolEqReducedPropagator}) with the definition
\begin{equation}
\label{eq:def:alpha_with_mu_g}
\begin{split}
\alpha_n(\tau) & = \sum_{\mu}|g_{n\mu}|^2 e^{-i \tilde{\omega}_{n\mu}\tau}.
\end{split}
\end{equation} 
Noting that  $  \alpha_n(\tau)  = \alpha_{1,n}(\tau) + \alpha_{2,n}(\tau) $ and using $\coth(\omega_{n\lambda}/2T) = 2\bar{n}_{n\lambda}+1$ one finds Eq.~(\ref{eq:def:alpha_with_lambda}).

\section{Functional expansion of auxiliary operators and Zeroth-Order Functional Expansion (ZOFE) approximation}
\label{sec:ZOFE-approach}

An often used approximation to treat the functional derivative of the general NMQSD equation is the so-called ZOFE approximation.
For completeness, we briefly show in this appendix how also in this approach the stochastics drops out of the equations and we present the final equations in the ZOFE treatment.

We follow \cite{DiGiSt98_1699_,StDiGi99_1801_,RoEiWo09_058301_,RoStEi11_034902_} and use the ansatz
\begin{equation}
\label{eq:def:Oop}
\frac{\delta}{\delta z_{s,n}^*}G(t,\boldsymbol{z}^*)\equiv  O_{n}(t,s,\boldsymbol{z}^*)G(t,\boldsymbol{z}^*),
\end{equation}
where we replace the action of the functional derivative on the reduced propagator $G(t,\boldsymbol{z}^*)$ in Eq.~(\ref{EvolEqCorrelatorWithNoise2}) by a linear operator $O_{n}(t,s,\boldsymbol{z}^*)$ in the electronic system that acts on $G(t,\boldsymbol{z}^*)$. 
Here, the lower index $n$ at the operator  $O_{n}(t,s,\boldsymbol{z}^*)$ indicates that this operator is connected to the coupling operator $K_n$ of monomer $n$.
Note that in  our previous work \cite{RoEiWo09_058301_,RoStEi10_5060_,RoStEi11_034902_,RiRoSt11_113034_,RiRoSt11_2912_} we used an upper index instead.

For convenience we define the auxiliary operators
\begin{equation}
\label{eq:def:Obar}
 \bar{O}_{n}(t,\boldsymbol{z}^*) \equiv \int_{0}^{t} ds\ \alpha_n(t-s) O_{n}(t,s,\boldsymbol{z}^*).
\end{equation}
Then Eq.~(\ref{EvolEqCorrelatorWithNoise2}) can be written as
\begin{equation}
\label{EvolEqPsiSysSpaceMitOQuer}
\begin{split}
 \partial_t \mathcal{C}(t) =
  -i H_{\rm sys} \mathcal{C}(t) -\sum_n K_n \int \frac{d^2 \boldsymbol{z}}{\pi}\,e^{-|z|^2} \\
  \times \bar{O}_{n}(t,\boldsymbol{z}^*) G(t,\boldsymbol{z}^*).
\end{split}
\end{equation}

The operators $\bar{O}_{n}(t, \boldsymbol{z}^*)$ in Eq.~(\ref{EvolEqPsiSysSpaceMitOQuer}) describe the coupling of monomer $n$ to its local environment and implicitly contain temperature through the environmental correlation function $\alpha_n(t-s)$ under the memory integral in Eq.~(\ref{eq:def:Obar}).
The evolution of ${O}_{n}(t, s,\boldsymbol{z}^*)$ can be obtained \cite{DiGiSt98_1699_,RoStEi11_034902_} from the consistency condition
\begin{equation}
\label{eq:consistency}
\frac{d}{dt} \big(O_{n}(t,s,\boldsymbol{z}^*) \psi(t,\boldsymbol{z}^*)\big) = \frac{\delta}{\delta z_{s,n}^*} \partial_t \ket{\psi(t,\boldsymbol{z}^*)}.
\end{equation}

We assume that $\bar{O}_{n}(t,\boldsymbol{z}^*)$ is analytic in $z_{t,n}^*$ and expand it in a Taylor series w.\,r.\,t.~the  time-dependent complex numbers (stochastic processes) $z_{t,n}^*$, which yields
\begin{align}
& \bar{O}_{n}(t,\boldsymbol{z}^*) = \bar{O}^{(0)}_{n}(t) \nonumber \\
& \quad + \sum_{n_{1}}\int_{0}^{t}\bar{O}^{(1),n_{1}}_{n}(t,v_{1})\, z^*_{v_{1},n_{1}}\, dv_{1} \nonumber \\
& \quad + \sum_{n_{1},n_{2}}\int_{0}^{t}\int_{0}^{t}\bar{O}^{(2),n_{1},n_{2}}_{n}(t,v_{1},v_{2})\,  z^*_{v_{1},n_{1}} z^*_{v_{2},n_{2}}\, dv_{1} dv_{2} \nonumber \\
& \quad + \ldots
\label{eq:_func_expansion_o-op}
\end{align}
with operator-valued expansion coefficients $\bar{O}^{(k),n_{1},\ldots,n_{k}}_{n}(t,v_{1},\ldots,v_{k})$. 
For these expansion coefficients one can derive coupled differential equations describing their time evolution \cite{YuDiGi99_91_}. When evaluating the integral in Eq.~(\ref{EvolEqPsiSysSpaceMitOQuer}) with the expansion Eq.~(\ref{eq:_func_expansion_o-op}) all terms with a $z_{t,n}^*$-dependence yield zero. This is because both $\bar{O}_n(t,\boldsymbol{z}^*)$ as well as $G(t,\boldsymbol{z}^*)$ are analytic in $z_{t,n}^*$ and therefore in the product $\bar{O}_{n}(t,\boldsymbol{z}^*) G(t,\boldsymbol{z}^*)$ only the combination of the two zeroth-order terms is left after the integration in Eq.~(\ref{EvolEqPsiSysSpaceMitOQuer}). The resulting equation for the correlation operator $\mathcal{C}(t)$ reads
\begin{equation}
\begin{split}
 \partial_t \mathcal{C}(t) = &\Big( -i H_{\rm sys} -\sum_n K_n \bar{O}^{(0)}_{n}(t)  \Big) \mathcal{C}(t),
\end{split}
\label{EvolEqCorrelatorWithoutNoise}
\end{equation}
with, indeed, only the $\boldsymbol{z}^*$-independent zeroth-order O-Operator $\bar{O}^{(0)}_{n}(t)$. Note that although only $\bar{O}^{(0)}_{n}(t)$ contributes to the evolution equation for $\mathcal{C}(t)$, still $\bar{O}^{(0)}_{n}(t)$ has to be determined by solving the coupled hierarchy of equations resulting from Eqs.~(\ref{eq:consistency}) and~(\ref{eq:_func_expansion_o-op}).

To obtain a numerically efficient method, previously we have applied an approximation~\cite{YuDiGi99_91_,RoEiWo09_058301_}, which we denote as `zeroth-order functional expansion' (ZOFE) approximation. For the case of absorption considered here this approximation amounts to neglecting the coupling of the zeroth order to the higher orders in the hierarchy.
The operators $O_{n}^{(0)}(t,s)$ are then determined by the auxiliary evolution equation
\begin{align}
\label{eq:O_0}
  \partial_t O^{(0)}_{n}(t,s)=&-i\Big[H_{\rm sys},O_{n}^{(0)}(t,s)\Big]\\
&+\sum_m\Big[\opP{m}\bar{O}_{m}^{(0)}(t),O_{n}^{(0)}(t,s)\Big]\nonumber
\end{align}
with the initial condition $O_{n}^{(0)}(s,s) = K_n = -\proj{\pi_n}$.

Note that the ZOFE approximation is exact in several important limiting cases, e.g., for weak electronic coupling between the monomers, for weak coupling to the environment, as well as in the Markov limit. Furthermore, in many parameter regimes relevant for molecular aggregates, the ZOFE approximation gives very precise results.

\section{Simplified formulas for the dimer}
\label{sec:Dimer}

We consider a dimer where the transition dipoles of the monomers are perpendicular to the distance vector $R_{12}$ between them and are rotated by an angle $\theta$ with respect to each other (see Fig.~\ref{fig:dimer_sketch}).
Then the expressions for $(A^{\rm Abs})_{nm}$ and $(A^{\rm CD})_{nm}$, Eqs.~(\ref{eq:A_abs}) and~(\ref{eq:A_CD}), can be easily evaluated. We obtain
\begin{align}
(A^{\rm Abs})_{11} = (A^{\rm Abs})_{22} & = \mu^2  \\
(A^{\rm Abs})_{12} = (A^{\rm Abs})_{21} & = \mu^2 \cos \theta
\end{align} 
and 
\begin{align}
(A^{\rm CD})_{11} = (A^{\rm CD})_{22} & = 0  \\
(A^{\rm CD})_{12} = (A^{\rm CD})_{21} & = -|R_{12}| \mu^2 \sin \theta.
\end{align} 
This leads to 
\begin{align}
\label{eq:_c_abs_12}
c^{\rm Abs}(t) &= \mu^2\big[ \big(\mathcal{C}_{11}(t)+\mathcal{C}_{22}(t)\big) + \cos\theta\big( \mathcal{C}_{12}(t) + \mathcal{C}_{21}(t)\big) \big], \\
c^{\rm CD}(t) &= -R\mu^2 \sin \theta \big[ \mathcal{C}_{12}(t) + \mathcal{C}_{21}(t) \big]
\label{eq:_c_cd_12}
\end{align}
with $R=|\vec{R}_{12}|$

Keeping in mind (see Subsection~\ref{sub:vector_propagation}) that the propagation of $\mathcal{C}_{nm}(t)$ is equivalent to propagating the initial state $\ket{\pi_m}$ until time $t$ and then projecting on the state $\ket{\pi_n}$ one can express the above Eqs.~(\ref{eq:_c_abs_12}) and~(\ref{eq:_c_cd_12}) in a way which is sometimes more convenient. For the propagation of the wave function one uses the same equations as for  $\mathcal{C}_{nm}(t)$ but with the matrix $\mathcal{C}_{nm}(t)$ replaced by a vector $\ket{\psi_{m}(t)}$, where the lower index $m$ indicates that the initial condition is $\ket{\psi_{m}(0)}=\ket{\pi_m}$.
With this one can write:
\begin{align}
c^{\rm Abs}(t)
  &= \mu^2\big[\cos^2(\theta/2) \big(\bra{\pi_1}+\bra{\pi_2}\big) \big(\ket{\psi_{1}(t)}+\ket{\psi_{2}(t)}\big) \nonumber \\
  &\qquad  + \sin^2(\theta/2) \big(\bra{\pi_1}-\bra{\pi_2}\big) \big(\ket{\psi_{1}(t)}-\ket{\psi_{2}(t)}\big)\big], \\
c^{\rm CD}(t) &= -R\mu^2 \sin \theta \big[ \braket{\pi_1}{\psi_{2}(t)}+ \braket{\pi_2}{\psi_{1}(t)} \big].
\end{align}
Note that $(\ket{\psi_{1}(t)}\pm\ket{\psi_{2}(t)}) / \sqrt{2} \equiv \ket{\psi^{\pm}(t)}$ is a state which can be obtained by propagating the initial state $\ket{\phi^{\pm}(t)}\equiv(\ket{\pi_1}\pm\ket{\pi_2})/\sqrt{2}$.

The above formulas can then be rewritten to
\begin{align}
\label{eq:_c_abs_plus_minus}
c^{\rm Abs}(t)
  &= 2 \mu^2 \big[ \cos^2(\theta/2) \braket{\phi_+}{\psi^+(t)} \nonumber \\
  &\qquad\quad     + \sin^2(\theta/2)\braket{\phi_-}{\psi^-(t)} \big] \\
\label{eq:_c_cd_plus_minus}
c^{\rm CD}(t)
  &= -4 R\mu^2 \sin \theta \big[ \braket{\phi_+}{\psi^+(t)} - \braket{\phi_-}{\psi^-(t)} \big].
\end{align}
Thus, the absorption spectrum can be calculated from a weighted sum of the 'symmetric' and 'antisymmetric' contributions.
The CD spectrum is just the difference of these two contributions with equal weight.

\bibliographystyle{journal_v5.bst}

\end{document}